\documentclass[aps,prb,twocolumn,superscriptaddress]{revtex4}

\usepackage{graphicx}% Include figure files
\usepackage{dcolumn}% Align table columns on decimal point
\usepackage{bm}% bold math
\usepackage{hyperref}
\begin{document}

\title{Hydrogenated Graphene Nanoribbons for Spintronics}

\date{\today}

\author{D. Soriano}
\affiliation{Departamento de F\'isica Aplicada, Universidad de Alicante, San Vicente del Raspeig, Alicante 03690, Spain}
\author{F. Mu\~noz-Rojas}
\affiliation{Departamento de F\'isica Aplicada, Universidad de Alicante, San Vicente del Raspeig, Alicante 03690, Spain}
\affiliation{Department of Physics, Korea Advanced Institute of Science and Technology, Daejeon 305-701, Korea}
\author{J. Fern\'andez-Rossier}
\affiliation{Departamento de F\'isica Aplicada, Universidad de Alicante, San Vicente del Raspeig, Alicante 03690, Spain}
\author{J. J. Palacios}
\affiliation{Departamento de F\'isica Aplicada, Universidad de Alicante, San Vicente del Raspeig, Alicante 03690, Spain}
\affiliation{Departamento de F\'isica de la Materia Condensada, Universidad Aut\'onoma de Madrid, Cantoblanco, Madrid 28049, Spain}

\begin{abstract}
%Recent proposals for graphene based spintronics are rooted on the magnetic properties of nanoribbons 
%or nanographenes with zigzag edges. 
%We propose in this work 
%an alternative to edge-related graphene spintronics which is based on exploiting bulk hydrogenation.
We show how hydrogenation of graphene nanoribbons at small concentrations 
can open new venues towards carbon-based spintronics 
applications regardless of any especific edge termination or passivation of the nanoribbons.   
Density functional theory calculations show that an adsorbed H atom induces a spin density on the 
surrounding $\pi$ orbitals whose symmetry and degree of localization
depends on the distance to the edges of the nanoribbon. As expected for graphene-based systems, 
these induced magnetic moments interact ferromagnetically or antiferromagnetically
depending on the relative adsorption graphene sublattice, but the magnitude of the 
interactions are found to strongly vary with
the position of the H atoms relative to the edges.  We also calculate, with the help of the Hubbard model,
the transport properties of 
hydrogenated armchair semiconducting graphene nanoribbons in the diluted regime and show
how the exchange coupling between H atoms can be exploited in the design of novel magnetoresistive devices.
 \end{abstract}


\maketitle

\section{introduction}
There is a widespread consensus on the large potetial of graphene for electronic
applications\cite{geim:35,Neto09}.
Theoretically, graphene also holds promise for a vast range of applications in spintronics,
although clear evidence of magnetic graphene is, however, elusive to date. 
In a broad sense, two factors may account for this elusiveness. First,  the fact that 
hydrocarbons of high spin
are known to be highly reactive and, unless fabricated or synthesized under very clean and
controled conditions, they will likely bind surrounding 
species with the concomittant disappearance of magnetism\cite{Palacio-book}.
A second reason relates to the fact that the ground state
of graphene is near an interaction-driven phase transition
into an insulating antiferromagnetic state\cite{Sorella-92}. 
This underlying antiferromagnetic correlations
prevent the magnetic moments, even if they develop, from ordering ferromagnetically and 
preclude the possibility of observing hysteresis in standard magnetic measurements.
Notwithstanding, a few reports of magnetic graphite\cite{PhysRevLett.91.227201}
and graphene\cite{Wang09-1} can be found in recent literature.
 
Most of recent theoretical ideas for graphene based spintronics applications are 
rooted on the magnetic properties of nanoribbons\cite{Fujita_1996,Son06,munoz-rojas:136810,Kim08-2,Gunlycke07} 
or nanographenes\cite{Harigaya02,fernandez-rossier:177204,wang:157201,Ezawa08-1} with zigzag edges. 
All these proposals assume a very particular edge hydrogenation where H atoms passivate 
the $\sigma$ dangling bonds, leaving
all the $\pi$ orbitals unsaturated and carrying the magnetic moments. 
However, this is just one out of many possible edge
realizations which range from H-free self-passivation\cite{koskinen:115502} to 
full H passivation\cite{wassmann:096402}. According to the work of Wassmann 
et al.\cite{wassmann:096402}, 
relatively low H concentations at room temperature suffice to completely
passivate the edges, including the edge $\pi$ orbitals responsible for the magnetic order. 
The self-passivated or reconstructed 
zigzag edges\cite{koskinen:115502} are, in fact, 
among the least energetically favorable of all, 
although, interestingly, have been recently observed by transmission electron microscopy\cite{Chuvilin09}.

\begin{figure}
\includegraphics[width=\linewidth]{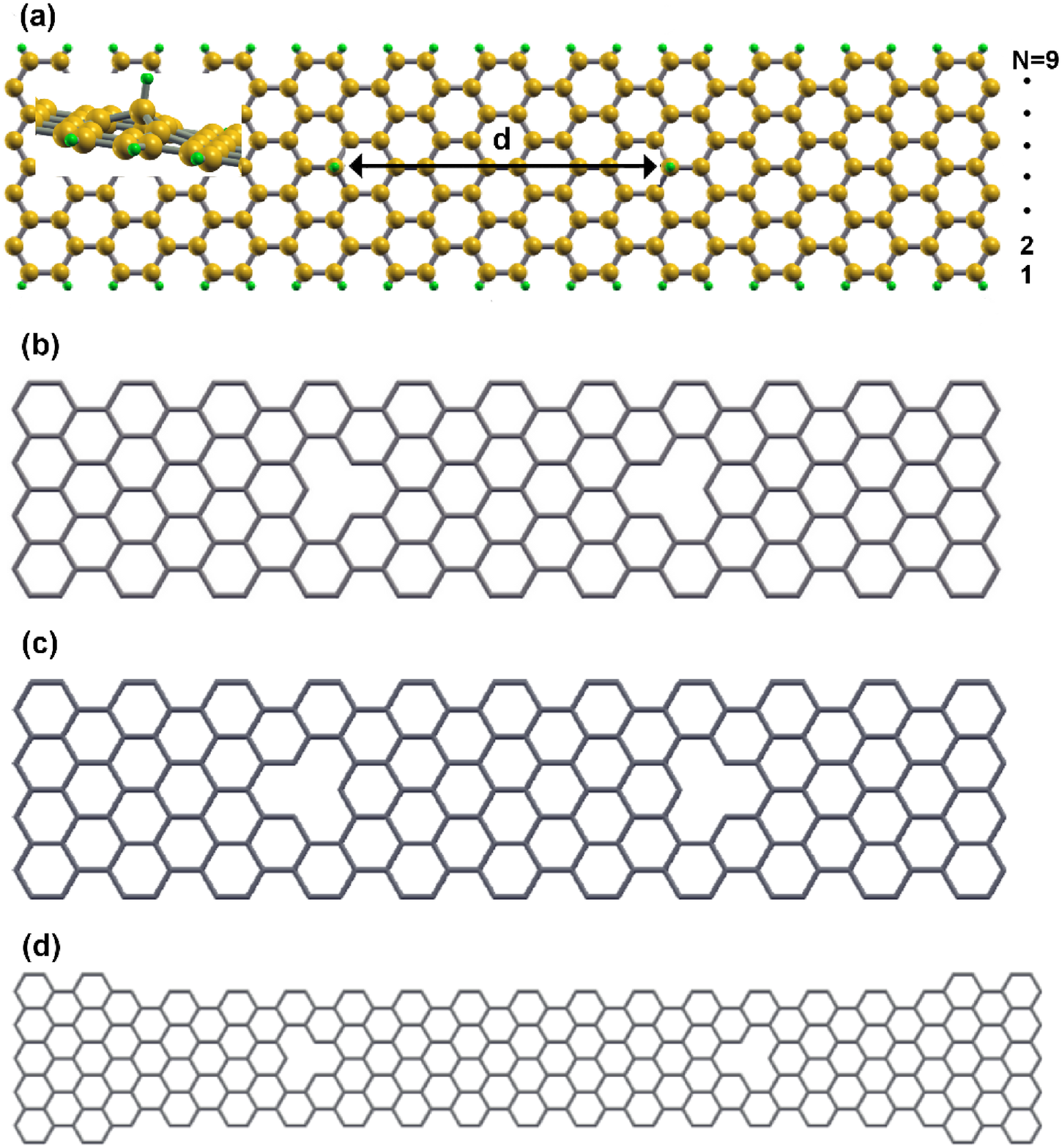}
  \caption{(a) Armchair graphene nanoribbon
of width $N$ (where $N$ is the number of dimer lines) with two H atoms (shown in green) 
adsorbed in the middle of the ribbon at a distance $d$ from each other. The inset shows 
a detail of the adsorption geometry. (b) Pictorical representation of the one-orbital tight-binding model 
where the presence of H atoms  is simulated by removing sites in a head-to-head configuration. 
(c) Same as in (b), but for the opposite ordering 
(tail-to-tail). (d) Same as in (b), but for a finite semiconducting AGNR connected to metallic nanoribbons
 at the edges. }
  \label{agnr}
\end{figure}

In the light of the present controversy on the actual possibilities of ever encounter zigzag magnetic edges, 
we propose in this work 
an alternative to edge-related spintronics in which to exploit the recently shown controled
hydrogenation of graphene\cite{D.C.Elias01302009}.
The key factor here is that adsorption of  atomic 
H in the bulk of graphene is accompanied by the appearance of a magnetic 
moment of 1$\mu_{\rm B}$ localized on the $\pi$ orbitals surrounding each H 
atom\cite{yazyev:125408}. These magnetic moments
interact with one another ferromagnetically or antiferromagnetically,
depending on whether their respective adsorption sublattices (usually labeled A and B)
are the same or not\cite{yazyev:125408,brey:116802,Casolo09}. 
Statistically speaking, a sublattice compensated H coverage is expected
unless adsorption on one sublattice is privileged, e.g., by the substrate.  To 
date, however, there is no evidence that such an uncompensated coverage can 
be achieved.  In the more likely compensated case an overall antiferromagnetic alignment with a total spin $ S =0$ 
is thus energetically favored over a ferromagnetic one with $S > 0$.
As a proof of principle and since we are interested in the diluted regime, we consider in this
work the fundamental problem of
two H atoms. These are covalently bonded to the surface of a semiconducting
armchair graphene nanoribbon (AGNR) on different sublattices [see Fig. \ref{agnr}(a)]. Here the $\sigma$ bonds of the 
edges are fully passivated with H so that they are irrelevant at low energies.

\begin{figure}
\includegraphics[width=\linewidth]{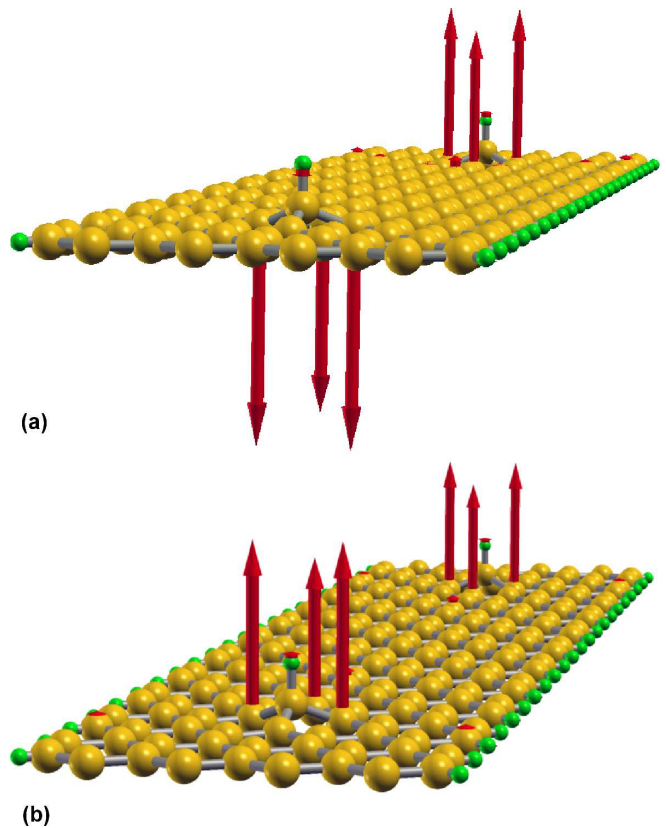}
  \caption{Pictorical view of the antiferromagnetic state (a) and the magnetic-field driven ferromagnetic state (b) 
where the magnetic moments localized around two hydrogen atoms are depicted by red arrows 
(the orientation of the arrows with respect to the graphene plane is arbitrary since spin-orbit coupling is neglected here).}
  \label{states}
\end{figure}

As shown in Sec. \ref{results}, after a brief introduction to the theoretical basics presented in Sec. \ref{theory},
the magnetic-field driven ferromagnetic (F) state, where the
H-induced magnetic moments are aligned by the field [as shown 
in Fig. \ref{states}(b)], can present a different resistance from that of 
the natural  antiferromagnetic (AF) zero-field state [as that in Fig. \ref{states}(a)]. 
Two different cases are discussed: Infinite semiconducting AGNR's and 
finite ones connected to conductive graphene [see schematic piture in Fig. \ref{agnr}(d)].
The differences in conductance between the F and the AF states 
can be substantial and translate into a magnetoresistive response as large as 100\% for 
distances between the H atoms of the order of few nanometers. Practical implications of these
results are  discussed in Sec. \ref{practical} and a brief summary is presented in Sec. \ref{summary} 

\section{Computational details}
\label{theory}
For the calculation of the electronic structure of hydrogenated AGNR's we use both
\textit{ab initio} techniques within the local spin density approximation (LSDA), 
aided by the CRYSTAL03 package\cite{CRYSTAL:03}, and
a one-orbital ($\pi$) first-neighbor tight-binding  model where the electronic repulsion is 
treated by means of a Hubbard-like interaction $U$ in the mean field approximation: 
\begin{equation}
\label{H-F}
 \hat H=\sum_{i,j}t\hat{c}^\dagger_i\hat{c}_j+ U\sum_i (\hat n_{i\uparrow}\langle \hat n_{i\downarrow}\rangle +
 \hat n_{i\downarrow}\langle \hat n_{i\uparrow}\rangle) - U\sum_i \langle \hat n_{i\downarrow}\rangle
 \langle \hat n_{i\uparrow}\rangle.
 \end{equation}
The first term represents the kinetic energy with first neighbors hopping $t$
between $\pi$ orbitals. The remaining terms account for the electronic interactions where
$\hat n_{i\eta}=\hat c^+_{i\eta}\hat c_{i\eta}$ are the number operators associated
to each $\pi$ orbital with spin $\eta$. 
These two different levels of approximation to the electronic structure
have been shown to yield similar results for nanoribbons\cite{fernandez-rossier:075430} and 
nanographenes\cite{fernandez-rossier:177204} where a full passivation by H of the $\sigma$ bonds is assumed.

The bulk adsorption geometry of a H atom has been obtained by relaxing the C atom bonded to the
H and the nearest C atoms until the characteristic $sp^3$ 
hybridization is obtained [see inset in Fig. \ref{agnr}(a)].  The bonding between a H atom and 
a C atom results in the effective removal of the
$\pi$ orbital from the low energy sector, so that the H adsorption is simulated
by simply removing a site in the one-orbital mean-field Hubbard model [see Fig. \ref{agnr}(b-d)]. 
Our results, as shown below, provide further support to the use
of the Hubbard model in graphene systems, as an alternative to the computationally more demanding LSDA and  
extend the range of applicability of Lieb's theorem\cite{PhysRevLett.62.1201} to a wider set of situations.

To calculate the transport properties we use the standard Green's function partitioning method
as implemented, e.g., in the quantum transport package ALACANT\cite{ALACANT:06}. 
To this purpose, the infinite system is divided into three parts, namely a central region (C), 
containing the H atoms, which is  
connected to the right (R) and left (L) semi-infinite clean leads. The Hamiltonian matrix 
describing the whole system is then given by
\begin{equation}
H=H_C+H_R+H_L+V_{LC}+V_{RC}
\label{eqn:1}
\end{equation}   
where $H_C$, $H_L$ and $H_R$ are the Hamiltonian matrices of the central region, the left and the right lead, respectively.
$V_{LC}$ and $V_{RC}$ represent the coupling between the central region and the leads. In general,
the non-orthogonality of the basis set
 must be taken into account when writing the Green's function of the central region:
\begin{equation}
{G}_C(E)=[ES_C-H_C-\Sigma_L(E)-{\Sigma}_R(E)]^{-1}
\label{eqn:2}.
\end{equation}
The self-energies of the left (${\Sigma}_L$) and right (${\Sigma}_R$) leads 
account for the influence of these on the electronic structure of the central part 
and $S_C$ is the overlap matrix in this region. 
%The total DOS can then be calculated by summing over 
%the diagonal elements of the Green's function
%
%\begin{equation}
%\rho_{\alpha}(E)=-\frac{1}{\pi}\sum_{\alpha}{\rm Im}({G}_C(E)S_C)_{\alpha,\alpha}
%\label{eqn:3}
%\end{equation}
%
For the calculation of the conductance we use the Landauer formula, $G(E)=\frac{e^2}{h} T(E)$. 
The transmission function $T$ can, in turn, be obtained from the expression:
\begin{equation}
T(E)=\sum_\eta{\rm Tr}\left[{G}_C^{\dagger}(E){\Gamma}_R(E){G}_C(E){\Gamma}_L(E)\right]_\eta,
\label{eqn:4}
\end{equation}
with ${\Gamma}_{R(L)}=i\left({\Sigma}_{R(L)}-{\Sigma}_{R(L)}^{\dagger}\right)$. Notice that since we are interested
in collinear magnetic solutions, all the matrices carry the spin index $\eta$, which we have not made explicit in previous
equations.  All the terms in Eq. \ref{eqn:1} must be obtained self-consistently either from a 
periodic boundary condition 
calculation, e.g., using CRYSTAL03  in the case of the LSDA calculations,
or following the methodology in Ref. \cite{munoz-rojas:136810} in the
case of the Hubbard model. The Fermi energy is set to zero and, in both cases, global
charge neutrality in all regions is imposed by shifting the onsite energies as necessary.

\begin{figure}
\includegraphics[width=\linewidth]{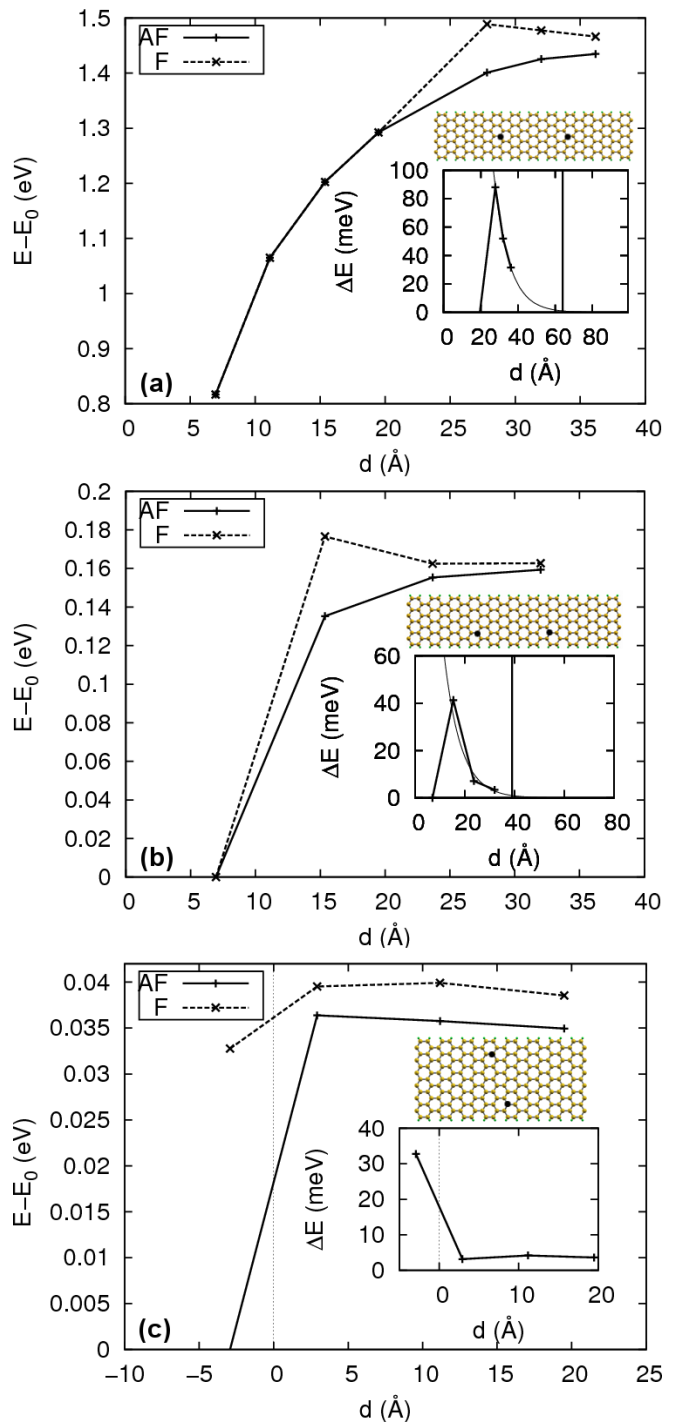}
  \caption{(a) Total energy referred to $E_0$ [lowest energy in case (b)] 
as a function of the distance between H atoms for the ferromagnetic (dashed)
and the antiferromagnetic (solid) state 
when placed in the middle of the ribbon. Upper inset: Picture of a nanoribbon with two H atoms. Lower inset:
Energy difference between both states and
extrapolation to large distances (solid line). The vertical line in the inset 
denotes the distance above which the energy difference becomes less than 1 meV.
  (b) Same as in (a), but for both H atoms near the same edge.
 (c) Same as in (a), but for H atoms placed on opposite edges. Here
the width of the ribbon is larger ($N=13$) and $E_0$ is the minimum energy among all the AF solutions.}
  \label{energetics}
\end{figure}

\section{results}
\label{results}
\subsection{Energetics}
We first examine the energetics of the F ($S=1$) and AF ($S=0$) states
as a function of the mutual distance $d$ between H atoms. We first choose a semiconducting AGNR of width $N=9 $, where
$N$ is the number of dimer lines across the ribbon,  and
restrict ourselves to the case of H atoms placed on different sublattices [see  Fig. \ref{agnr}(a)]. 
The reason for this choice is three-fold. The AB (or BA) 
configurations are always energetically preferred to the AA or BB configurations for similar 
distances between H atoms\cite{Casolo09}. Second, and
most importantly for the purpose of this work, the magnetic state of the AB (or BA)
configuration can be tuned by a magnetic field. Furthermore, as briefly mentioned in the introduction, even if
energetic considerations are left aside, the AB (or BA) 
configuration represents the simplest case of a random ensemble of H atoms which, in average, will
equally populate the two graphene sublattices. 

\begin{figure}
\includegraphics[width=\linewidth]{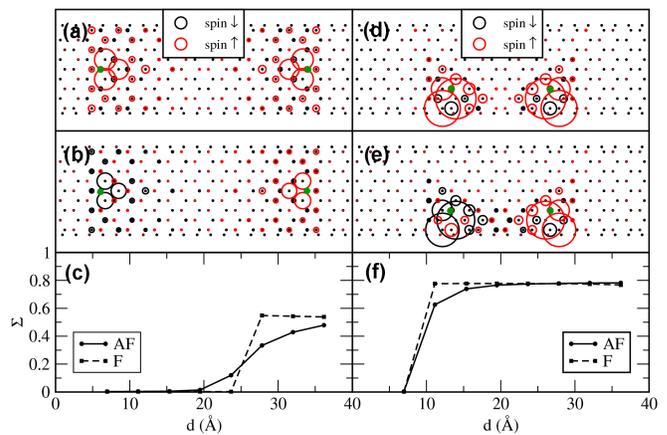}
  \caption{Magnetic moments (whose magnitude is represented by the size of the circles)
on individual C atoms when the H atoms are placed head-to-head in the middle of an  $N=9$ armchair graphene 
nanoribbon at a distance
$d=32.0$ \AA $\:$[(a) ferromagnetic state and (b) antierromagnetic state]
 and at a distance $d=15.4$ \AA$\:$ close to the edge 
[(d) ferromagnetic state and (e) antiferromagnetic state]. Panels (c) and (f) 
show the magnetization (see definition of $\Sigma$ in text) as a function of the distance between H atoms. }
  \label{mag}
\end{figure}

In  Fig. \ref{energetics}(a) we plot the LSDA total
energy for H atoms in the middle of the AGNR as a function
of $d$. The $S=0$ state is always 
the ground state. This implies antiferromagnetic coupling, except at short distances for 
which  the local magnetization, quantified through
$\Sigma =\sqrt{\sum_i \langle m_i \rangle^2}$, 
vanishes altoghether [see Fig. \ref{mag}(c)]. The quenching of the magnetization is 
easily understood in terms of the formation of a 
spin singlet\cite{palacios:195428}. At the minimum distance for which $\Sigma \ne 0$, the energy 
difference presents a maximum and decays exponentially for larger $d$ [see inset in  Fig. \ref{energetics}(a)]. 
As $d\rightarrow \infty$
the spin clouds do not interact anymore and both F and AF solutions tend to
have the same energy.     In summary, for any distance between H atoms
the ground state presents $S=0$, 
following Lieb's theorem\cite{PhysRevLett.62.1201}, but the overall spin texture strongly 
depends on their mutual distance. We note that it also depends on the ordering (AB or BA) of the H atoms. 
Whereas in bulk the spin cloud associated to the H atom would be invariant under rotations of 120 degrees, 
in a nanoribbon there is a preferential direction along the ribbon axis which is different for H atoms located on 
A and B sublattices [see Fig. \ref{agnr}(b-c)].  We refer to the preferential direction as the head
and the tail to the opposite one.  Thus, a head-to-head coupling  (AB) is expected to be 
much stronger than a tail-to-tail coupling (BA). 
The magnetization clouds are shown for an AB (or head-to-head) 
case in Fig. \ref{mag}(a-b). It is easy to appreciate the strong directionality just alluded to.
As a consequence, when reversing the ordering of the H atoms to
a BA (or tail-to-tail) configuration [see Fig. \ref{agnr}(c)], 
these do not couple magnetically at any distance, except when in very close proximity for which $\Sigma=0$.  
All these results are similar to the ones reported using the one-orbital mean-field Hubbard model\cite{palacios:195428}. 

In Fig. \ref{energetics}(b) we present the LSDA total
energy of the F and AF states when the H atoms are placed near one of the edges. Again the
ground state is the AF state for any distance. The proximity to the edge decreases the 
localization length of the spin texture, increasing $\Sigma$ [see Fig. \ref{mag}(d-e)], and 
thereby, decreasing the critical distance below which the magnetization disappears [see Fig. \ref{mag}(f)]. 
The perturbation of the edge modifies the spin texture around each atom and,
contrary to the previous case, when reversing the ordering to a tail-to-tail configuration, 
the H atoms couple magnetically in a finite range of distances (see below).
All energies, including those depicted in Fig. \ref{energetics}(a),
are referred to $E_0$, the lowest energy solution
for H atoms close to the edge, corresponding to the smallest distance. 
Note that for the same distance between H atoms
the total energy is always smaller when these are closer to the edge, which reflects that the binding energy 
is higher there by approximately 1 eV\cite{Boukhvalov08-1}. 

Finally we present in  Fig. \ref{energetics}(c) 
the case where the H atoms are placed on opposite edges for an $N=13$  semiconducting AGNR.
Placing the atoms on different edges allows us now to 
explore the energetics of different magnetic coupling orientations (tail-to-tail for $d>0$ and head-to-head for $d<0$) 
at a reasonable computational cost ($d$ is now the longitudinal distance between H atoms).
As expected, due to the strong anisotropy of the spin
texture, the magnetic coupling strongly depends on the ordering of the atoms and not only on their mutual
distance.  While for $d<0$ the energy difference 
between the AF and F states is large, for $d>0$ this difference decreases substantially and does not depend
too much on $d$ for the values considered. This asymmetry is easily understood since for $d>0$($d<0$) the spin textures
approximately couple in a tail-to-tail(head-to-head) manner. 
As shown below, this may have important experimental consequences.

\subsection{Magnetoresistance}
\begin{figure}
\includegraphics[width=\linewidth]{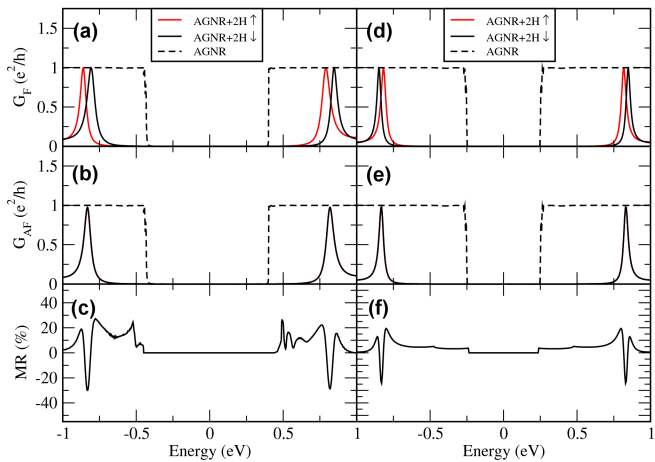}
  \caption{Spin resolved transmission as a function of energy for the ferromagnetic state
of an armchair graphene nanoribbon of width $N=9$  with two H atoms adsorbed in the 
middle at a mutual distance $d=32$ \AA $\:$ calculated in the (a)  local spin density approximation
 and (d) with a one-orbital 
mean-field Hubbard model. (b) and (e) panels show 
the same, but for the antiferromagnetic state. The resulting magnetoresistance in both approximations 
is shown in (c) and (f).  }
  \label{mr}
\end{figure}

We now turn our discussion to the implications these results may have on the electrical transport.
Under the influence of a magnetic field, the hydrogenated AGNR behaves like a diluted paramagnetic 
semiconductor\cite{lin:6554} for small
concentrations of H. At large concentrations, when the spin density is zero everywhere, the influence of the field can
only give rise to a minor diamagnetic response.
At intermediate concentrations, where the magnetization clouds induced by the H atoms interact with each other, one can 
switch from the AF to the F state by applying a sufficiently strong magnetic field.  
In analogy with the H$_2$ molecule, where
switching from the singlet to the triplet state modifies the orbital part of the wavefunction, here the 
electronic structure will be indirectly affected by the field even if its direct influence
on the orbital wave function is neglected.  This change reflects in the spin-resolved
conductance as shown in Fig. \ref{mr}(a-b) for $d=32$  \AA $\:$(the dashed line represents the conductance
of the clean AGNR). The different 
total transmission for the F and AF solutions, resulting from adding the two spin channels,
results in a positive magnetoresistance (MR),  $MR=G_{F}-G_{AF}/G_{F}+G_{AF}$, at energies near the 
bottom and top of the conduction and valence bands, respectively [see Fig. \ref{mr}(c)]. Similar results are
obtained (not shown) for different  intermediate distances between H atoms.
The right panels in Fig. \ref{mr} show the results obtained from the one-orbital mean-field Hubbard
model for $U=t=3$ eV.  Apart from the recovery of the 
particle-hole symmetry, the results are remarkably similar, 
validating the use of the latter model for transport calculations in hydrogenated graphene.

One should note, however, that since the chemical potential lies in the middle of the gap,
the energy ranges at which  MR could manifest itself are not relevant in linear response transport for infinite AGNR's.
A finite bias calculation may reveal the MR obtained at those energies, but this is a non-trivial task 
beyond the scope of this work.  The application of a gate voltage is not a practical alternative either since it
implies a deviation from charge neutrality which would fill up or empty the 
localized states hosting the unpaired spins and kill the magnetization. Instead, we propose
to explore the possibility of MR at zero bias by considering {\em finite} AGNR's
connected at the ends to conductive graphene. This is done in 
our calculations by considering a metallic AGNR with a
narrower section in the middle of appropriate width [see Fig. \ref{agnr}(d)]. 
(Note that in the one-orbital mean-field Hubbard approximation AGNR's
of width $N=3m-1$, where $m$ is an integer,  are metallic, being semiconducting otherwise). 
In what follows and in the light of the previous results, we restrict ourselves to the one-orbital 
mean-field Hubbard model.
\begin{figure}
\includegraphics[width=\linewidth]{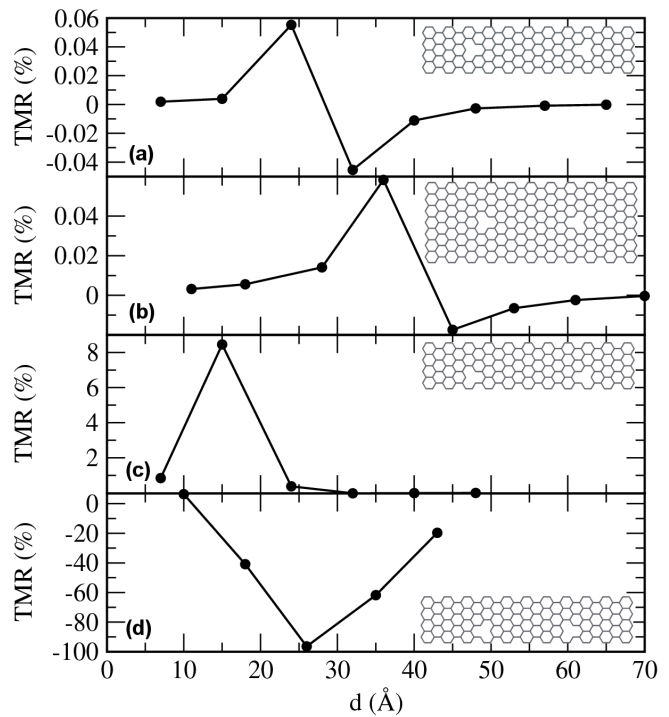}
  \caption{Tunneling magnetoresistance for four different atomic H configurations. Two 
H atoms in a head-to-head configuration located in the middle of an armchair graphene nanoribbon of
length $L=73.7$ \AA $\:$ with a width of $N=9$ (a) and $N= 15$ (b), 
and two H atoms located near one edge in a head-to-head (c) 
and in a tail-to-tail (d) configuration for a ribbon width of $N=9$ and length of $L=57.0 $ \AA.
 }
  \label{tmr}
\end{figure}

\begin{figure}
\includegraphics[width=\linewidth]{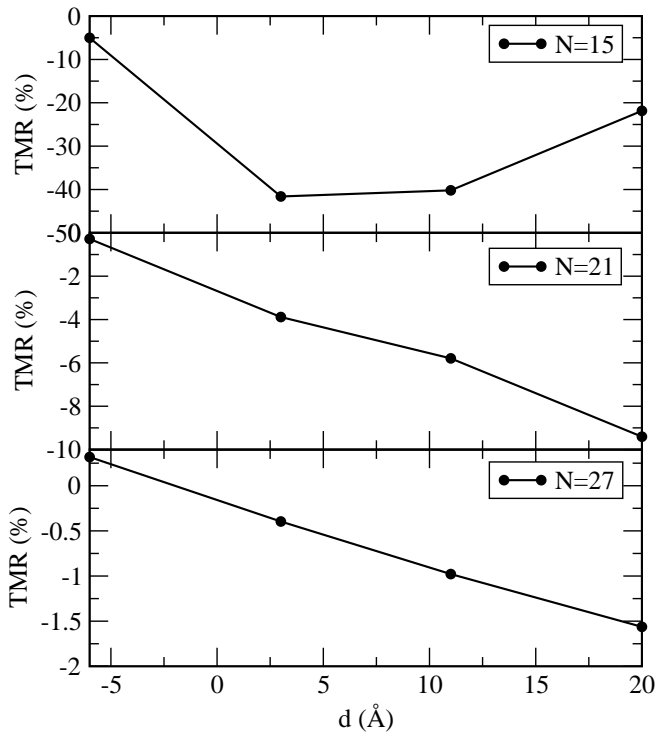}
  \caption{Tunneling magnetoresistance
 as a function of the longitudinal distance between H atoms for three different armchair graphene nanoribbon
 widths. The H atoms
are located on opposite edges of the ribbon. }
  \label{tmr-across}
\end{figure}

In our proposed AGNR heterostructure 
the difference in the zero-bias tunnel conductance between the F and AF
states is now responsible for the appearance of {\em tunneling} MR (TMR), as shown in 
Fig. \ref{tmr}. Notice that unlike conventional TMR, where the magnetic elements are in the electrodes, 
in our proposal magnetism is in the barrier.  Panels (a) and (b) in Fig. \ref{tmr}
correspond to H atoms placed in the middle of a semiconducting AGNR of length $L=73.7$ \AA $\:$
and width $N=9$ and $N=15$, respectively. Both cases refer to head-to-head configurations.
The obtained TMR changes sign with $d$, but it is always negligibly small.
On the contrary, when placed near the edge [Fig. \ref{tmr}(c)], the TMR is positive and reaches much larger values. This
result is for an $N=9$, $L=57.0 $ \AA $\:$ AGNR.
When the H atoms are now placed in a tail-to-tail configuration near the edge for the same 
AGNR, the TMR becomes negative 
and reaches values of up to 100\% [see Fig. \ref{tmr}(d)]. As a final example we show in Fig. \ref{tmr-across} the TMR for the 
case where the H atoms are placed on the opposite edges of an $L=57.0$ \AA $\:$ AGNR for three different widths.
 Large (and negative) values are also obtained for the tail-to-tail configurations in the range of distances explored. 
As can be appreciated, the TMR logically decreases with the ribbon width.

\section{Practical considerations}
\label{practical}
A critical assesment of the practical consequences of the results presented is due at this point:
\begin{itemize}
\item First, the TMR results presented above have been obtained for a particular type of AGNR heterostructure
and rely on the existence of metallic AGNR's. Calculations where the relaxation of the 
atomic structure on the edges is considered reveal that a gap always opens even for the 
nominally metallic armchair nanoribbons\cite{son:216803}, compromising the applicability of these nanoribbons
as electrodes. Graphene-based metallic leads, however, can be found in recent literature. 
For instance, zigzag graphene nanoribbons with passivated edges are metallic\cite{wassmann:096402}.
The physics described in this work does not rely on the edge termination of the nanoribbons and 
could be reproduced in fully passivated zigzag nanoribbons as well.
Another alternative can be based on using partially unzipped metallic 
carbon nanotubes\cite{Kosynkin}, as suggested in Ref. \cite{santos:086801}. 
A third possibility consists in gating selectively a semiconducting AGNR as previously done for 
nanotubes\cite{Javey:nl:04:a}. 
\item From the LSDA calculations we note that the energy difference between the F and AF states, or exchange coupling
energy, can be as large as tens of meV for short distances, particularly for head-to-head configurations. 
However, as far as magnetoresistive properties is concerned,
exchange coupling energies above 1 meV are of no
practical use since the Zeeman energy gain per spin in a magnetic field $B$ 
is $0.067 g\mu_B B$ meV $T^{-1}$ and fields higher than 10 T are hardly accesible in the lab.
The AF-F transition is therefore practically forbidden below $d\approx65$ \AA $\:$ and $d\approx40$ \AA$\:$  
in the examples shown in Fig. \ref{energetics}(a) and (b), respectively. 
At larger distances the fields necessary to induce the AF-F transition can be
as small as needed, but also is the associated MR as shown in Fig. \ref{tmr}(a) and (c). 
This is not a problem, however, when the H atoms are coupled tail-to-tail as, e.g., in the cases where they are located
on opposite edges. There, as shown in Fig. \ref{tmr-across}, a TMR as high as $\approx 50\%$ can be obtained at reasonably 
small exchange coupling energies [see inset in Fig. \ref{energetics}(c)]. 
\item The third caveat relates to the stability of the atomic configurations.
Given the tendency for H atoms to form lowest-energy non-magnetic 
aggregates\cite{Boukhvalov08-1}, the existence of magnetically active 
dilute ensembles of adsorbed H atoms requires certain conditions.
For instance, working at reasonable low temperatures prevents   
H atoms from difusing after adsorption\cite{Herrero09}. 
Another possibility is one intrinsic to AGNR's: The binding 
energy is larger for H atoms close to the edge than in the bulk of the nanoribbon. 
Related to this is the fact that 
the mobility of H decreases significantly near the edge, reducing the possiblity
of formation of lowest-energy non-magnetic aggregates near the edge\cite{Boukhvalov08-1}. 
As we have shown, it is precisely the
hydrogenation near the edge that favors the appearance of sizable MR and thus an actual experimental verification.  
\end{itemize}

\section{summary}
\label{summary}
In summary, we have shown, as a proof of principle, that hydrogenated AGNR's at small concentrations
can exhibit magnetoresistive properties without invoking purely edge-related physics. Our results, which 
are deeply rooted in Lieb's theorem, provide further evidence for the wide applicability
of this theorem beyond the bipartite lattice Hubbard model 
for which was demonstrated. We have also shown that hydrogenation near
the edge presents advantages with respect to bulk hydrogenation both from energetic and electronic standpoints. This 
aspect, especific to nanoribbons, might favor the use of these ones for 
spintronics applications, as opposed to using large flakes of graphene. 
Although our conclusions are based on the simplest case of two H atoms, nothing
seems to prevent magnetoresistance from occuring for emsembles of many H atoms. This, however,
still needs to be supported by a more extensive statistical analysis which is beyond the scope of this work.

\begin{acknowledgments}
This work has been financially supported by MICINN of Spain (Grants MAT07-67845 and CONSOLIDER CSD2007-00010). 
D.S. acknowledges financial support from Consejo Superior de Investigaciones Cient\' ificas (CSIC).
\end{acknowledgments}

\end{document}